\documentstyle[sprocl,epsf]{article}

\makeatletter

\def\alt{\mathrel{\mathpalette\vereq<}}
\def\vereq#1#2{\lower3pt\vbox{\baselineskip1.5pt \lineskip1.5pt
\ialign{$\m@th#1\hfill##\hfil$\crcr#2\crcr\sim\crcr}}}
\def\agt{\mathrel{\mathpalette\vereq>}}

\newcount\@tempcntc
\def\@citex[#1]#2{\if@filesw\immediate\write\@auxout
{\string\citation{#2}}\fi
  \@tempcnta\z@\@tempcntb\m@ne\def\@citea{}\@cite{\@for\@citeb:=#2\do
    {\@ifundefined
       {b@\@citeb}{\@citeo\@tempcntb\m@ne\@citea
        \def\@citea{,\penalty\@m\ }{\bf ?}\@warning
       {Citation `\@citeb' on page \thepage \space undefined}}%
{\setbox\z@\hbox{\global\@tempcntc0\csname b@\@citeb\endcsname\relax}%
     \ifnum\@tempcntc=\z@ \@citeo\@tempcntb\m@ne
       \@citea\def\@citea{,\penalty\@m}
       \hbox{\csname b@\@citeb\endcsname}%
     \else
      \advance\@tempcntb\@ne
      \ifnum\@tempcntb=\@tempcntc
      \else\advance\@tempcntb\m@ne\@citeo
      \@tempcnta\@tempcntc\@tempcntb\@tempcntc\fi\fi}}\@citeo}{#1}}

\def\@citeo{\ifnum\@tempcnta>\@tempcntb\else\@citea
  \def\@citea{,\penalty\@m}%
  \ifnum\@tempcnta=\@tempcntb\the\@tempcnta\else
   {\advance\@tempcnta\@ne\ifnum\@tempcnta=\@tempcntb \else
\def\@citea{-}\fi
    \advance\@tempcnta\m@ne\the\@tempcnta\@citea\the\@tempcntb}\fi\fi}

\makeatother

\begin{document}


\title{\vbox to0pt
{\vskip0pt minus100cm\centerline{\fbox{\parbox{\textwidth}
{\small\noindent To be published in
{\it Proceedings 
1998 Summer School in High-Energy Physics and Cosmology,  
ICTP, Trieste, Italy, 29 June--17 July 1998},
ed.\ by G.~Senjanovi\'c and A.Yu.~Smirnov (World Scientific,
Singapore).}}}
\vskip1.5cm}\vskip-22pt\noindent
NEUTRINO ASTROPHYSICS AT THE CROSS ROADS}
\author{G.G.~RAFFELT}
\address{Max-Planck-Institut f\"ur Physik 
(Werner-Heisenberg-Institut)\\
F\"ohringer Ring 6, 80805 M\"unchen, Germany}

\maketitle

\abstracts{Nonstandard neutrino properties (masses, mixing, sterile
  states, electromagnetic interactions, and so forth) can have
  far-reaching ramifications in astrophysics and cosmology.  We look
  at the most interesting cases in the light of the powerful current
  indications for neutrino oscillations.}


\section{Introduction}

The indications for neutrino oscillations from the atmospheric and
solar neutrino anomalies and from the LSND experiment are now so
overwhelming that the discourse in neutrino physics has changed.  One
no longer asks if these particles indeed oscillate, one rather debates
the most plausible pattern of masses and mixing angles, and if the
existence of a sterile neutrino is required.  Of course, all of the
indications for oscillations are to various degrees preliminary, yet
so intruiging that it is difficult to resist their charm.  

It is a truism that astrophysics and cosmology play a unique role in
neutrino physics, and conversely, that these light, weakly interacting
particles are absolutely crucial for some of the most interesting
astrophysical phenomena such as core-collapse supernovae and for the
universe at large. Therefore, in this brief survey of current topics
in neutrino astrophysics it behoves us to discuss what astrophysics
and cosmology contribute to the current debate in neutrino physics and
what the future perspectives are.

To this end we begin in Sec.~\ref{sec:neutrinooscillations} with an
overview of the current indications for neutrino oscillations and
possible global interpretations.  Astrophysical neutrinos, i.e.\ those
from the Sun and from cosmic-ray interactions in the upper atmosphere,
play a dominant role in this context.  In Sec.~\ref{sec:cosmology} we
next turn to the cosmological arguments relevant to neutrino physics
(dark matter, structure formation, cosmic microwave background,
big-bang nucleosynthesis).  Supernova (SN) neutrinos are the topic of
Sec.~\ref{sec:supernova} where we discuss the role of neutrino masses
and oscillations in this environment, the interpretation of the
SN~1987A neutrino burst, and what one could learn from a future
galactic SN.  The recent developments in high-energy neutrino
astronomy are touched upon in Sec.~\ref{sec:neutrinoastronomy}, while
in Sec.~\ref{sec:electromagneticproperties} 
astrophysical aspects of 
neutrino electromagnetic
properties are briefly discussed in the light of the current evidence
for oscillations. Finally, in Sec.~\ref{sec:conclusions} we summarize
our conclusions.


\section{Evidence for Neutrino Oscillations}
\label{sec:neutrinooscillations}

\subsection{Atmospheric Neutrinos}

The current evidence for neutrino oscillations arises from the
atmospheric neutrino anomaly, the solar neutrino problem, and the LSND
experiment.  It is probably fair to say that at present the most
convincing indication comes from atmospheric neutrinos.  We thus begin
our short survey with this spectacular case that has changed the
perception of this field.

The Earth is immersed in a diffuse flux of high-energy cosmic rays
consisting of protons and nuclei. The upper atmosphere acts as a
``beam dump'' where these particles quickly lose their energy by the
production of secondary pions (and some kaons) which subsequently
decay according to the simple scheme
\begin{eqnarray}\label{eq:beamdump}
\pi^+\to\mu^++\nu_\mu,\qquad \mu^+\to e^++\nu_e+\bar\nu_\mu,
\nonumber\\
\pi^-\to\mu^-+\bar\nu_\mu,\qquad \mu^-\to e^-+\bar\nu_e+\nu_\mu.
\end{eqnarray} 
The expected unequal flavor distribution
$\nu_e:\nu_\mu:\nu_\tau\approx 1:2:0$ allows one to use the
atmospheric neutrino flux to search for flavor oscillations.  Of
course, at energies beyond a few GeV the muons do not all decay before
hitting the Earth so that the $\nu_\mu/\nu_e$ flavor ratio increases
with energy.  Still, while the absolute neutrino flux predictions have
large uncertainties, perhaps on the 20\% level, the expected flavor
ratio is thought to be nearly model independent and calculable for all
relevant energies to within a few
percent~\cite{Gaisser96,Honda95,Agrawal96}.

First events from atmospheric neutrinos were measured in two
pioneering experiments in the mid-sixties~\cite{Achar65,Reines65}, but
it is only since the late eighties that several large underground
detectors began to address the question of flavor oscillations in
earnest~\cite{Nusex89,Frejus89,Frejus95,Kamiokande88,%
Kamiokande94,Kamiokande98,IMB91,IMB92,Soudan99,Macro98,SuperK98a,%
SuperK98b}.  Around 1988 the Kamiokande water Cherenkov detector
revealed a significantly reduced $\nu_\mu/\nu_e$ flavor ratio---the
atmospheric neutrino anomaly~\cite{Kamiokande88}. There was no
alternative explanation to oscillations, but a ``smoking-gun''
signature became available only with the high counting rates of
SuperKamiokande~\cite{SuperK98a,SuperK98b} which has taken data since
April~1996.

For a given solid angle, the atmospheric neutrino flux from above
should be equal to that produced in the atmosphere of the antipodes
because the $r^{-2}$ flux dilution with distance cancels a
corresponding increase in surface area. However, SuperKamiokande
observed a pronounced up-down-asymmetry in the multi-GeV sample
(visible energy deposition in the detector exceeding 1.33~GeV).  Using
the zenith-angle range $-1.0\leq\cos\theta\leq-0.2$ as defining
``up,'' and the corresponding range $0.2\leq\cos\theta\leq1.0$ for
``down,'' the $\nu_e+\bar\nu_e$ flux shows a ratio~\cite{Kajita98}
${\rm up/down}=0.93^{+0.13}_{-0.12}$ while $\nu_\mu+\bar\nu_\mu$ has
$0.54^{+0.06}_{-0.05}$. It is this up-down-asymmetry which gives one
confidence that there is no simple explanation in terms of the
neutrino production process in the atmosphere or the experimental
flavor identification.

\begin{figure}[b]
\epsfxsize=5.8cm \hbox to\hsize{\hss\epsfbox{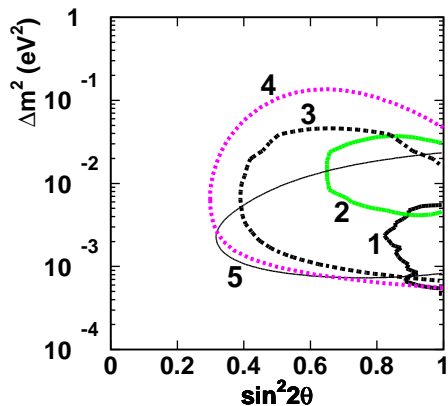}\hss}
\caption{Allowed mixing parameters at 90\% CL from atmospheric
  neutrinos for $\nu_\mu\to\nu_\tau$
  oscillations~\protect\cite{Kajita98}.  They are based on the
  contained events in SuperKamiokande~(1) and Kamiokande~(2), the
  upward through-going muons in SuperKamiokande~(3) and
  Kamiokande~(4), and the stopping fraction of upward going muons in
  SuperKamiokande~(5). (Figure reproduced with kind permission of
  T.~Kajita.)
\label{fig:atmo1}}
\end{figure}

Neutrino oscillations, on the other hand, provide a simple and
consistent interpretation.  In the usual two-flavor formalism with a
vacuum mixing angle $\Theta$, the appearance probability for the
oscillation from a flavor $\nu$ to $\nu'$ is
\begin{equation}\label{eq:oscillation}
P(\nu\to\nu')=\sin^22\Theta\,\,
\sin^2\left(1.27\,\frac{\Delta m_\nu^2}{\rm eV^2}\,
\frac{L}{{\rm km}}\,\frac{{\rm GeV}}{E_\nu}\right).
\end{equation}
If the $\nu_\mu$'s oscillate into $\nu_\tau$'s with a nearly maximal
mixing angle and if $\Delta m_\nu^2$ is of order $10^{-3}~{\rm eV}^2$,
one obtains the observed behavior since the relevant energies are a
few GeV and $L$ to the other side of the Earth is around $10^4~{\rm
km}$.  The detailed 90\% CL contours for the allowed range of mixing
parameters from different signatures in Kamiokande and SuperKamiokande
are summarized in Fig.~\ref{fig:atmo1}.  Meanwhile, more data have
been taken, shifting the curve~(1) to somewhat larger $\Delta m_\nu^2$
values~\cite{DPF99atmo} with the boundaries shown in
Table~\ref{tab:osci}.

\begin{figure}[ht]
\epsfxsize=7cm \hbox to\hsize{\hss\epsfbox{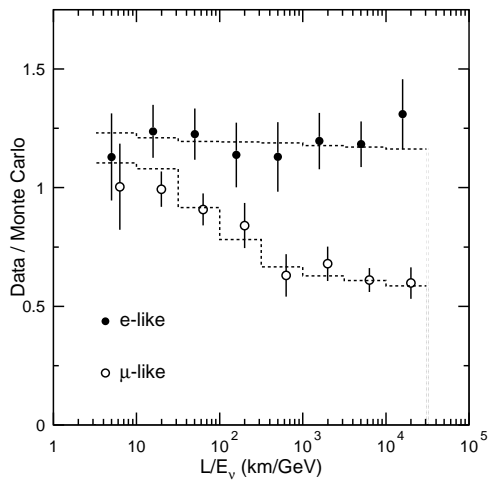}\hss}
\caption{$L/E_\nu$ plot for the fully contained events at 
SuperKamiokande~\protect\cite{SuperK98b}. The points
show the ratio of measured counts over Monte Carlo expectation
in the absence of oscillations. The dashed lines show the
expectation for $\nu_\mu\to\nu_\tau$ oscillations with
$\Delta m_\nu^2=2.2\times10^{-3}~{\rm eV^2}$ and
$\sin^22\Theta=1$. (Figure reproduced with kind permission of
  T.~Kajita.)
\label{fig:atmo2}}
\end{figure}

Equation~(\ref{eq:oscillation}) suggests that one should plot the data
according to their $L/E_\nu$ as in Fig.~\ref{fig:atmo2}.  This
representation provides perhaps the most convincing argument for the
reality of atmospheric neutrino oscillations.  The flat distribution
of the $\nu_e$ points excludes $\nu_\mu\to\nu_e$ oscillations as a
dominant channel, in agreement with the CHOOZ limits on this
mode~\cite{Chooz98}.  Therefore, $\nu_\mu\to\nu_\tau$ or oscillations
into a sterile channel $\nu_\mu\to\nu_s$ are favored.

A calculation of the $\nu_\mu\to\nu_s$ oscillation probability 
must include the refrac\-tive energy shift in the Earth.  Recall that
the neutrino weak potential~is
\begin{equation}
V_{\rm weak}=\pm\,\frac{G_F n_B}{2\sqrt2}\times
\cases{-2Y_n+4Y_e&for $\nu_e$,\cr
-2Y_n&for $\nu_{\mu,\tau}$,\cr
0&for $\nu_s$,\cr}
\end{equation}
where the upper sign refers to neutrinos, the lower sign to
antineutrinos, $G_F$ is the Fermi constant, $n_B$ the baryon density,
$Y_n$ the neutron and $Y_e$ the electron number per baryon (both about
1/2 in normal matter). Numerically we have
\begin{equation}
\frac{G_F n_B}{2\sqrt2}=
1.9\times10^{-14}~{\rm eV}~\frac{\rho}{\rm g~cm^{-3}}.
\end{equation}
The dispersion relation is $E_\nu=V_{\rm weak}+
\sqrt{p_\nu^2+m_\nu^2}$ so that $V_{\rm weak}$ should be compared with
$m_\nu^2/2p_\nu$. For $\Delta m_\nu^2$ around $10^{-3}~{\rm eV}^2$,
$p_\nu$ of a few GeV, and $\rho$ of a few $\rm g~cm^{-3}$, the energy
difference between $\nu_\mu$ and $\nu_s$ arising from $V_{\rm weak}$
is about the same as that from $\Delta m_\nu^2/2p_\nu$.  The resulting
modification of the oscillation pattern can cause rather peculiar
zenith-angle distributions~\cite{Liu98a,Liu98b,Lipari98}, but the
current data do not allow one to exclude the $\nu_s$ channel.

While the $\nu_\tau$ is quasi-sterile in the detector because of the
large mass of the $\tau$-lepton, there is still an important
difference to a $\nu_s$ because the $\nu_\tau$ produces pions in
neutral-current collisions such as $\nu N\to N \nu \pi^0$ which can be
seen by $\pi^0\to2\gamma$. With better statistics and a dedicated
analysis one may be able to distinguish the $\nu_\tau$ and $\nu_s$
oscillation channels~\cite{Vissani98,Learned98,Hall98}.

The evidence for atmospheric neutrino oscillations is very compelling,
yet an independent confirmation is urgently needed.  Hopefully it will
come from one of the long-baseline experiments where an accelerator
neutrino beam is directed toward a distant detector.  The most
advanced project is the K2K experiment~\cite{K2K} between KEK and
Kamioka with a baseline of 250~km.  Other projects include detectors
in the Soudan mine at a distance of 730~km from
Fermilab~\cite{Minos,EmulsionSandwich}, or in the Gran Sasso
Laboratory at 732~km from CERN~\cite{Icarus,Noe,Opera}.

\begin{figure}[b]
\epsfxsize=10cm \hbox to\hsize{\hss\epsfbox{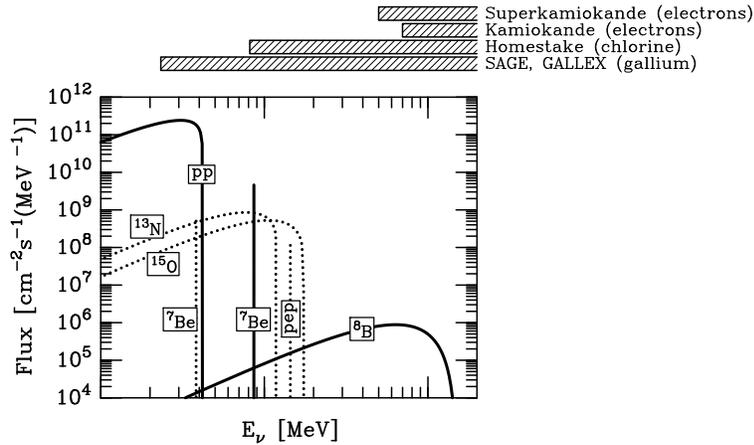}\hss}
\caption{Solar neutrino flux at Earth. Continuum spectra in ${\rm
cm^{-2}~s^{-1}~MeV^{-1}}$, line spectra in ${\rm
cm^{-2}~s^{-1}}$. Solid lines are the sources of dominant experimental
significance.
\label{fig:sunspectrum}}
\end{figure}

\subsection{Solar Neutrinos}

The Sun, like other hydrogen-burning stars, liberates nuclear binding
energy by the effective fusion reaction $4p+2e^-\to{}^4{\rm
He}+2\nu_e+ 26.73~{\rm MeV}$ so that its luminosity implies a $\nu_e$
flux at Earth of $6.6\times10^{10}~{\rm cm^{-2}~s^{-1}}$. In detail,
the production of helium involves primarily the pp-chains---the CNO
cycle is important in stars more massive than the Sun.  The expected
solar neutrino flux is shown in Fig.~\ref{fig:sunspectrum}, where
solid lines are for the three contributions which are most important
for the measurements,
\begin{eqnarray}
\hbox to 1.7cm{pp:\hfil}
&\hbox to5.0cm{$p+p\to{}^2{\rm H}+e^++\nu_e$\hfil}
&(E_\nu<0.420~{\rm MeV}),
\nonumber\\
\hbox to 1.7cm{Beryllium:\hfil}
&\hbox to5.0cm{$e^-+{}^7{\rm Be}\to{}^7{\rm Li}+\nu_e$\hfil}
&(E_\nu=0.862~{\rm MeV}),
\nonumber\\
\hbox to 1.7cm{Boron:\hfil}
&\hbox to5.0cm{$p+{}^7{\rm Be}\to{}^8{\rm B}
\to{}^8{\rm Be}^*+e^++\nu_e$\hfil}
&(E_\nu\alt15~{\rm MeV}).
\end{eqnarray}
A crucial feature of these reactions is that the beryllium and boron
neutrinos both arise from ${}^7$Be which may either capture a proton
or an electron so that their relative fluxes depend on the branching
ratio between the two reactions. 

\begin{figure}[b]
\epsfxsize=9cm\hbox to\hsize{\hss\epsfbox{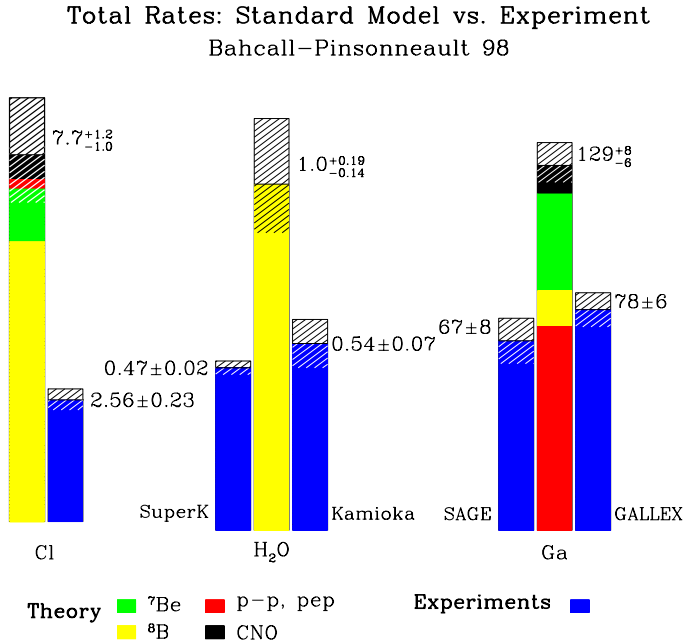}\hss}
\caption{Solar neutrino fluxes measured in five experiments vs.\
theoretical predictions from a standard solar
model~\protect\cite{BP98}. (Figure courtesy of J.~Bahcall.)
\label{fig:sunproblem}}
\end{figure}

The solar neutrino flux has been measured in five different
experiments with three different spectral response characteristics;
the relevant energy range is indicated by the hatched bars above
Fig.~\ref{fig:sunspectrum}. The radiochemical gallium experiments
GALLEX~\cite{Gallex99a,Gallex99b} and SAGE~\cite{Sage} reach to the
lowest energies and pick up fluxes from all source reactions. The
Homestake chlorine experiment~\cite{Homestake} picks up beryllium and
boron neutrinos, while the Kamiokande~\cite{Kamiokande} and
SuperKamiokande~\cite{SuperKsun,DPF99solar} water Cherenkov detectors
see only the upper part of the boron flux. All of the experiments see
a flux deficit relative to standard-solar model
predictions as summarized in Fig.~\ref{fig:sunproblem} and in 
a recent overview~\cite{Bahcall98a}.

It has been widely discussed that there is no possibility to account
for the measured fluxes by any apparent astrophysical or
nuclear-physics modification of the standard solar models so that an
explanation in terms of neutrino oscillations is difficult to
avoid~\cite{Bahcall98a,Castellani97}. Moreover, at something like the
99.8\% CL one cannot account for the measurements by an
energy-independent global suppression factor~\cite{Bahcall98a}.
Therefore, one cannot appeal to neutrino oscillations with an
arbitrary $\Delta m_\nu^2$ and a large mixing angle.

One viable possibility are vacuum oscillations with a large mixing
angle and a $\Delta m_\nu^2$ around $10^{-10}~{\rm eV}^2$, providing
an oscillations length of order the Sun-Earth distance and thus an
energy-dependent suppression factor. Second, one can have solutions
with $\Delta m_\nu^2$ in the neighborhood of $10^{-5}~{\rm eV}^2$
where the mass difference between the oscillating flavors (energy of
order 1~MeV) can be canceled by the neutrino refractive effect in the
Sun, leading to resonant or MSW oscillations~\cite{Mikheyev85,Kuo89},
again with an energy-dependent suppression factor of the $\nu_e$
flux. In this case one may have a nearly maximal mixing angle, or a
small one as shown in Table~\ref{tab:osci}. The large-angle MSW region
does not provide a credible fit for $\nu_e\to\nu_s$ oscillations while
the other solutions are possible for the $\nu_e\to\nu_{\mu,\tau}$ or
$\nu_e\to\nu_s$ channels, of course with somewhat different contours
of preferred mixing parameters~\cite{Bahcall98a}. 
It is noteworthy that the spectral distortion of the spectrum of
recoil electrons measured at SuperKamiokande seems to single out the
vacuum case as the preferred solution~\cite{DPF99solar}, although this
must be considered a rather preliminary conclusion at present.

\subsection{LSND}

The LSND (Liquid Scintillation Neutrino Detector) experiment is 
the only case of a pure laboratory experiment which shows 
indications for neutrino oscillations~\cite{LSND96}. 
It utilizes a proton beam at the Los Alamos National
Laboratory in the US.  The protons are directed at a target
where neutrinos arise from the same basic mechanism
Eq.~(\ref{eq:beamdump}), upper line, that produces them in the
atmosphere.  From $\pi^+$ decay-in-flight one obtains a $\nu_\mu$ beam
of up to 180~MeV while the subsequent decay-at-rest of stopped
$\mu^+$'s provides a $\bar\nu_\mu$ beam of less than 53~MeV.  The beam
should not contain any $\bar\nu_e$'s; they can be detected by
$\bar\nu_e p\to n e^+$ in coincidence with $np\to d\gamma$(2.2~MeV).
For energies above 36~MeV, the 1993--95 data included 22 such events
above an expected background of $4.6\pm0.6$; this excess is
interpreted as evidence for $\bar\nu_\mu\to\bar\nu_e$ oscillations.

The LSND data favor a large range of $\nu_e$-$\nu_\mu$-mixing
parameters. After taking the exclusion regions of other experiments
into account, one is left with a sliver of mixing parameters in the
range indicated in Table~\ref{tab:osci}.  The KARMEN experiment is
also sensitive in this range, but has not seen any
events~\cite{KARMEN98}. This lack of confirmation, however, does not
exclude the LSND evidence as the non-observation of only a few
expected events is not a statistically persuasive conflict.  Moreover,
if one excludes the background-infested 20--36~MeV data in LSND one
finds a much broader range of allowed mixing parameters than could
have been probed by KARMEN~\cite{Caldwell98b}.  Within 2--3 years all
of the LSND area will be covered with high sensitivity by
MiniBooNE~\cite{Boone}, a new experiment at Fermilab, which will
settle this case.

\begin{table}[b]
\caption{Experimentally favored neutrino mass differences and
mixing angles.\label{tab:osci}}
\smallskip
\hbox to\hsize{\hss\vbox{\hbox{\begin{tabular}[4]{llll}
\hline\noalign{\vskip2pt}\hline\noalign{\vskip2pt}
Experiment&Favored Channel&$\Delta m^2$ [$\,\rm eV^2$]
&$\sin^22\Theta$\\
\noalign{\vskip2pt}\hline\noalign{\vskip2pt}
LSND&$\bar\nu_\mu\to\bar\nu_e$&0.2--10&(0.2--$3)\times10^{-2}$\\
Atmospheric&$\nu_\mu\to\nu_\tau$&(1--8)${}\times10^{-3}$&0.85--1\\
&$\nu_\mu\to\nu_s$&(2--7)${}\times10^{-3}$&0.85--1\\
Solar\\
\quad Vacuum&$\nu_e\to{}$anything&$(0.5$--$8)\times10^{-10}$&0.5--1\\ 
\quad MSW (small angle)&$\nu_e\to{}$anything&(0.4--1)${}\times10^{-5}$
&$10^{-3}$--$10^{-2}$\\
\quad MSW (large angle)
&$\nu_e\to\nu_\mu$ or $\nu_\tau$&
(3--30)${}\times10^{-5}$&0.6--1\\
\noalign{\vskip2pt}
\hline
\noalign{\vskip2pt}
\hline
\end{tabular}}}\hss}
\end{table}

\subsection{Global Interpretation}

In Table~\ref{tab:osci} we summarize the neutrino oscillation channels
and mixing parameters indicated by the atmospheric and solar neutrino
anomalies and the LSND experiment. Clearly there is no straightforward
interpretation because there are too many indications!  If only three
different mass eigenstates $m_i$, $i=1,2,3$, exist, the mass
splittings must satisfy
\begin{equation}
\sum_{\rm Splittings} \Delta m_\nu^2
=(m_3^2-m_2^2)+(m_2^2-m_1^2)+(m_1^2-m_3^2)=0,
\end{equation}
a trivial condition which is not met by the independent $\Delta
m_\nu^2$ from Table~\ref{tab:osci}.  Some of the experiments may not
be due to a single $\Delta m_\nu^2$ but rather to nontrivial
three-flavor oscillation
patterns~\cite{Acker97,Cardall97,Teshima98,Thun98}.  Even then it
appears that one must ignore some of the experimental evidence or
stretch the errors beyond plausible limits to accommodate all
experiments in a three-flavor scheme.

\begin{figure}[ht]
\epsfxsize=7cm \hbox to\hsize{\hss\epsfbox{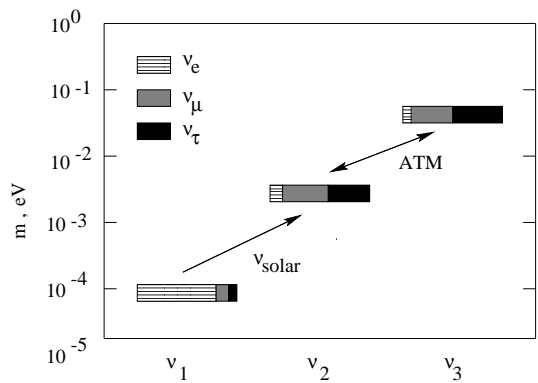}\hss}
\caption{Hierarchical mass and mixing scheme to account for solar
and atmospheric neutrinos, the former by the small-angle MSW
solution. The flavor content of each mass eigenstate is indicated by
the fill-patterns.  (Figure~\protect\cite{Smirnov99}
reproduced with kind permission of A.~Smirnov.)
\label{fig:mass1}}
\end{figure}

If one has to throw out one of the indications, LSND is usually taken
as the natural victim because there is no independent confirmation,
and because the other cases simply look too strong to be struck from
the list.  Once LSND has been disposed of, a typical mass and mixing
scheme may be as shown in Fig.~\ref{fig:mass1} where the small-angle
MSW solution has been taken for solar neutrinos.

However, the large mixing angle which is needed to account for the
atmospheric neutrino anomaly suggests that more than one mixing angle
may be large. Moreover, the spectral distortion observed in
SuperKamiokande suggests that the solar vacuum solution may be
preferred~\cite{DPF99solar}. Of course, the vastly different values
for $\Delta m_\nu^2$ implied by atmospheric neutrinos and the solar
vacuum solution looks unnatural.  Shrugging off this objection, there
are several workable schemes involving more than one large mixing
angle, for example bi-maximal mixing or threefold maximal
mixing~\cite{Smirnov99}.

It is also conceivable that the mass differences are not
representative of the masses themselves, i.e.\ that all three flavors
have, say, an eV-mass with small splittings as implied by solar and
atmospheric neutrinos (degenerate mass pattern). Of course, such a
scheme is very different from the hierarchical patterns that we know
in the quark and charged-lepton sectors, but the large mixing angle or
angles look very unfamiliar, too. If the neutrino masses are all
Majorana, one may still evade bounds on the effective $\nu_e$ Majorana
mass $\langle m_{\nu_e}^2\rangle_{\rm eff}$ relevant for neutrinoless
$\beta\beta$ decay. For example, in the bi-maximal mixing case there
is an exact cancellation so that $\langle m_{\nu_e}^2\rangle_{\rm
eff}=0$ in the limit where the mass differences can be neglected
relative to the common mass scale.

\begin{figure}[ht]
\epsfxsize=7cm \hbox to\hsize{\hss\epsfbox{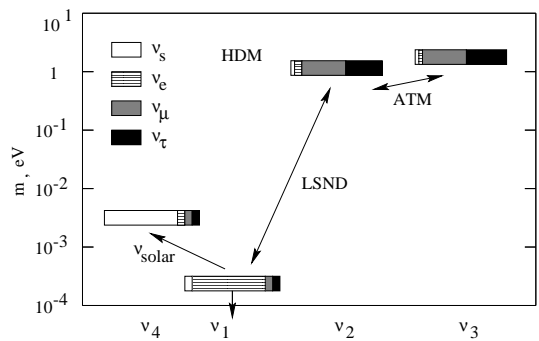}\hss}
\caption{Representative four-flavor mass and mixing scheme to account
for all experimental evidence.  (Figure~\protect\cite{Smirnov99}
reproduced with kind permission of A.~Smirnov.)
\label{fig:mass2}}
\end{figure}

At the present time there is no objective reason to ignore LSND.  As a
consequence, a very radical conclusion follows: there must be four
independent mass eigenstates, i.e.\ at least one low-mass
neutrino degree of freedom beyond the three sequential flavors. This
fourth flavor $\nu_s$ would have to be sterile with regard to the
standard weak interactions.  Probably the most natural mass and mixing
pattern is one like Fig.~\ref{fig:mass2}, but there are also other
possibilities~\cite{Smirnov99,Valle98}.

Of course, it would be an extremely radical and unexpected finding if
the oscillation experiments had not only turned up evidence for
neutrino masses, but for an additional, previously unsuspected
low-mass sterile neutrino. A confirmation of LSND by
MiniBooNE~\cite{Boone} would make this conclusion difficult to avoid
so that this new experiment is perhaps the most urgent current effort
in experimental neutrino physics.


\section{Cosmology}
\label{sec:cosmology}

\subsection{Big-Bang Nucleosynthesis}

Massive neutrinos and the existence of sterile neutrinos can have a
variety of important cosmological consequences. One immediately
wonders if a fourth neutrino flavor is not in conflict with the
well-known big-bang nucleosynthesis (BBN) limit on the effective
number of thermally excited primordial neutrino degrees of
freedom~\cite{Malaney93,Sarkar96,Schramm98}.  However, there are
several questions. The first and most obvious one is whether the
observationally inferred light-element abundances strictly exclude a
fourth flavor at the epoch of BBN.  The unfortunate answer is that,
while a fourth flavor clearly would make a very significant
difference, BBN is not in a position to exclude this possibility with
the sort of confidence that would be required to dismiss the
sterile-neutrino hypothesis~\cite{Olive98}.

Second, a sterile neutrino need not attain thermal equilibrium in the
first place.  It is excited by oscillations in conjunction with
collisions so that its contribution to the cosmic energy density at
the BBN epoch depends on the mass difference and mixing angle with an
active flavor~\cite{Barbieri91b,Cline92,Enqvist92,Shi93,Cardall96,%
Kirilova97,Bilenky98}. If the atmospheric neutrino anomaly is due to
$\nu_\mu\to\nu_s$ oscillations, the large mixing angle and large
$\Delta m_\nu^2$ imply that the sterile neutrino would be fully
excited at the time of BBN. On the other hand, for the small-angle MSW
solution or the vacuum solution of the solar neutrino problem, it is
barely excited so that the additional energy density is
negligible. Therefore, of the different four-flavor patterns BBN
favors those where $\nu_e$-$\nu_s$ oscillations solve the solar
neutrino problem over those where $\nu_\mu$-$\nu_s$ oscillations
explain the atmospheric neutrino anomaly.

Even this conclusion can be avoided if a lepton asymmetry of order
$10^{-5}$ exists at the time of the primordial $\nu_\mu\to\nu_s$
oscillations~\cite{Foot95}. It may be possible to create such
asymmetries among the active neutrinos by oscillations between, say,
$\nu_\tau$ ($\bar\nu_\tau$) and sterile states~\cite{Shi96,Foot97},
although the exact requirements on the mass and mixing parameters are
controversial in some
cases~\cite{Bell98,Foot98,Shi98a,Foot98b,Shi98b}.

Be that as it may, a sterile neutrino provides for a rich oscillation
phenomenology in the early universe, but at the same time BBN is not
quite enough of a precision tool to distinguish seriously between
different four-flavor patterns.  As it stands, BBN would benefit more
from pinning down the neutrino mass and mixing pattern experimentally
than the other way round.

\subsection{Dark Matter}

Irrespective of the possible existence of a sterile neutrino, it has
become difficult to dispute that neutrinos have masses. Therefore,
they could play an important role for the cosmological dark matter.
Standard calculations in the framework of the big-bang cosmology
reveal that the present-day universe contains about $100~{\rm
cm^{-3}}$ neutrinos and antineutrinos per active flavor~\cite{Kolb90},
leading to a cosmological mass fraction of
\begin{equation}
\Omega_\nu h^2=\sum_{i=1}^3\frac{m_i}{93~{\rm eV}},
\end{equation} 
where $h$ is the Hubble constant in units of $100~{\rm
km~s^{-1}~Mpc^{-1}}$.  The observed age of the universe together with
the measured expansion rate reveals that $\Omega h^2\alt0.4$, leading
to the most restrictive limit on the masses of all neutrino
flavors~\cite{Gershtein66,Cowsik72}. Once we believe the current
indications for oscillations, the mass differences are so small that
this limit reads $m_\nu\alt 13~{\rm eV}$ for the common mass scale of
all flavors, roughly identical with the world-averaged tritium
endpoint limit on $m_{\nu_e}$ of about~\cite{Caso98} 15~eV.

If the neutrino masses were in this range they could be the cosmic
dark matter as first pointed out more than 25 years
ago~\cite{Cowsik73}. However, it was quickly recognized that neutrinos
do not make for a good universal dark matter candidate. The simplest
counter-argument (``Tremaine-Gunn-limit'')
arises from the phase space of spiral galaxies which
cannot accommodate enough neutrinos to account for their dark matter
unless the neutrino mass obeys a {\it lower}
limit~\cite{Tremaine79,Madsen91}.  For typical spiral galaxies it
is~\cite{Salucci97} $m_\nu\agt20~{\rm eV}$, for dwarf galaxies even
$m_\nu\agt100$--200~eV, difficult to reconcile with the cosmological
upper limit.

\subsection{Large-Scale Structure}

The Tremaine-Gunn-limit is only the tip of the iceberg of evidence
against neutrino dark matter. The most powerful argument arises from
cosmic structure formation. At early times the universe was extremely
smooth as demonstrated by the tiny amplitude of the temperature
fluctuations of the cosmic microwave background radiation across the
sky.  The present-day distribution of matter, on the other hand, is
very clumpy. There are stars, galaxies, clusters of galaxies, and
large-scale coherent structures on scales up to about $100~{\rm Mpc}$.
A perfectly homogeneous expanding universe stays that way forever.
The standard theory~\cite{Kolb90,Boerner92,Coles95,Primack97} for the
formation of structure has it that the universe was initially almost,
but not quite, perfectly homogeneous, with a tiny modulation of its
density field. The action of gravity enhances the density contrast as
time goes on, leading to the observed structures.

The outcome of this evolution depends on the initial spectrum of
density fluctuations which is 
usually taken to be approximately flat, i.e.\ of the
``Harrison-Zeldovich-type,'' corresponding to the power-law-index
$n=1$. However, the {\it effective\/} spectrum relevant for structure
formation is the processed spectrum which obtains at the epoch when
the universe becomes matter dominated.  As the matter which makes up
the cosmic fluid can diffuse around, the smallest-scale density
fluctuations will be wiped out. This effect is particularly important
for weakly interacting particles which can diffuse far while they are
relativistic. Low-mass particles stay relativistic for a long time and
thus wipe out the primordial fluctuations up to large scales.  Massive
particles stay put earlier and thus have this effect only on small
scales. One speaks of ``hot dark matter'' (HDM) if the particle masses
are small enough that all fluctuations are wiped out beyond scales
which later correspond to a galaxy. Conversely, ``cold dark matter''
(CDM) has this effect only on sub-galactic scales.

\begin{figure}[b]
\epsfxsize=8cm \hbox to\hsize{\hss\epsfbox{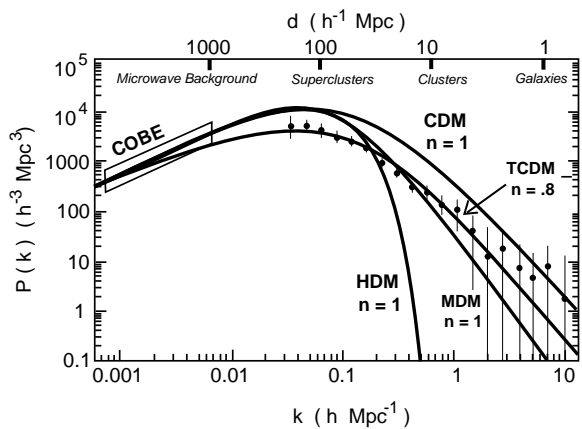}\hss}
\caption{Comparison of matter-density power spectra for cold dark
matter (CDM), tilted cold dark matter (TCDM), hot dark matter (HDM),
and mixed hot plus cold dark matter (MDM) for large-scale structure
formation~\protect\cite{Steinhardt95}. All curves are
normalized to COBE and include only linear approximation; nonlinear
corrections become important on scales below about
$10\,\rm Mpc$.
(Figure reproduced with kind permission of P.~Steinhardt.)
\label{fig:struc}}
\end{figure}

One way of presenting the results of calculations of structure
formation is to show the expected power-spectrum of the present-day
matter distribution (Fig.~\ref{fig:struc}) which can be compared to
the observed galaxy distribution. The theory of structure formation
then predicts the form, but not the amplitude of the spectrum which
can be fit either on large scales to the observed temperature
fluctuations of the cosmic microwave background radiation as observed
by the COBE satellite, or else on small scales to the observed galaxy
distribution. Figure~\ref{fig:struc} illustrates that HDM (neutrinos)
suppresses essentially all small-scale structure below a cut-off
corresponding to a supercluster scale and thus does not seem to be
able to account for the observations.

While cold dark matter works impressively well, it has the problem of
producing too much clustering on small scales. Ways out include a
primordial power spectrum which is not strictly flat (tilted dark
matter), a mix of cold and hot dark matter, or the assumption of a
cosmological constant.  Currently there is a broad consensus that some
variant of a CDM cosmology where structure forms by gravitational
instability from a primordial density fluctuations of approximately
the Harrison-Zeldovich type is probably how our universe works.

Thus, while it is widely accepted that neutrinos are not the main
dark-matter component, quite conceivably they contribute something
like 20\%, giving rise to a hot plus cold dark matter (HCDM) scenario
which avoids the overproduction of small-scale structure of a pure CDM
cosmology~\cite{Primack95,Pogosyan95,Klypin97,Gross98}.  A HDM
fraction exceeding about 20\% is inconsistent with the size of voids
in the galaxy distribution~\cite{Ghigna94}.  It was claimed that the
HCDM picture with about 20\% HDM provides the best fit to all current
large-scale structure data~\cite{Gross98,Gawiser98}.  Moreover, if
LSND is confirmed, especially with a $\Delta m_\nu^2$ of around
$6~{\rm eV}^2$, there would be a cosmic HDM component of just the
right magnitude~\cite{Primack95}. The LSND signal and the HCDM
cosmologies have become closely intertwined issues.

\begin{figure}[b]
\epsfxsize=6cm \hbox to\hsize{\hss\epsfbox{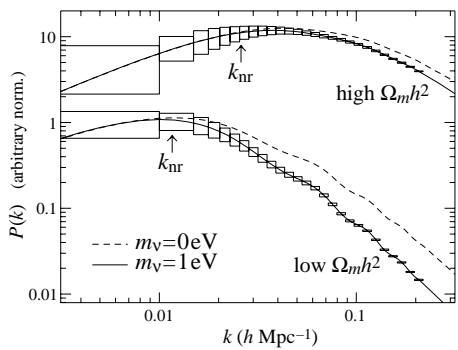}\hss}
\caption{Effect of a 1~eV neutrino mass on the power spectrum of the
distribution of bright-red galaxies compared with the expected
$1\,\sigma$ sensitivity of the Sloan Digital Sky Survey (error
boxes)~\protect\cite{Hu98}.
Upper curves: $\Omega_M=1$, $h=0.5$ with or without
a neutrino mass. Lower curves:  $\Omega_M=0.2$, $h=0.65$.
(Figure reproduced with kind permission of M.~Tegmark)
\label{fig:sloan}}
\end{figure}

However, important arguments against a HCDM scenario have appeared.
First, a cosmological model with the critical amount of dark matter is
hard to reconcile with all the evidence on the matter density;
something like 30\% looks far more convincing.  Moreover, the
high-redshif type~Ia supernova Hubble diagram now indicates the
existence of a cosmological constant
$\Lambda$~\cite{Perlmutter98,Garnavich98,Riess98,Schmidt98}.  If
correct, one is naturally led to a critical cosmological model with
something like 5\% baryonic matter, 25\% CDM, and 70\% ``vacuum
energy.''  Likewise, the observed abundance of high-redshift
($z\sim3$) galaxies is reproduced in this type of $\Lambda$CDM model,
but not by HCDM~\cite{Somerville98}.  A small amount of HDM is still
possible in a $\Lambda$CDM scenario, but not especially needed for
anything~\cite{Primack98}.

The cosmic large-scale structure is sensitive to small neutrino
masses, whether or not they are needed.  Put another way, the unknown
common mass scale which is left open by oscillation experiments has a
measurable impact on the power spectrum of the large-scale matter
distribution. For example, the upcoming Sloan Digital Sky
Survey~\cite{Sloan} will produce precision data where a neutrino mass
as small as 0.1~eV makes a noticeable difference~\cite{Hu98}, even
though a statistically meaningful neutrino mass limit may not lie far
below 1~eV. This is illustrated in Fig.~\ref{fig:sloan} where the
expected Sloan sensitivity to the power spectrum of bright red
galaxies is compared with theoretical predictions in a universe with
the critical mass in dark matter ($\Omega_M=1$) and a low-density
universe ($\Omega_M=0.2$), each time with or without a 1~eV neutrino.

In the long-term future, weak lensing of galaxies
by large-scale structure may provide even more precise information on
cosmological parameters. An ultimate sensitivity to a neutrino mass as
low as 0.1~eV has been suggested~\cite{Hu98b}.

\subsection{Cosmic Microwave Background Radiation}

Another sensitive probe of large-scale structure is the cosmic
microwave background radiation (CMBR), and more specifically the
power-spectrum of its temperature fluctuations across the sky. The
anticipated sky maps of the future MAP~\cite{MAP} and
PLANCK~\cite{PLANCK} satellite missions have already received advance
praise as the ``Cosmic Rosetta Stone''~\cite{Bennett97} because of the
wealth of cosmological precision information they are expected to
reveal~\cite{White94,Jungman96b,Bond97,Hu98c}.

\begin{figure}[ht]
\hbox to\hsize{\hss\epsfxsize=6.7cm\epsfbox{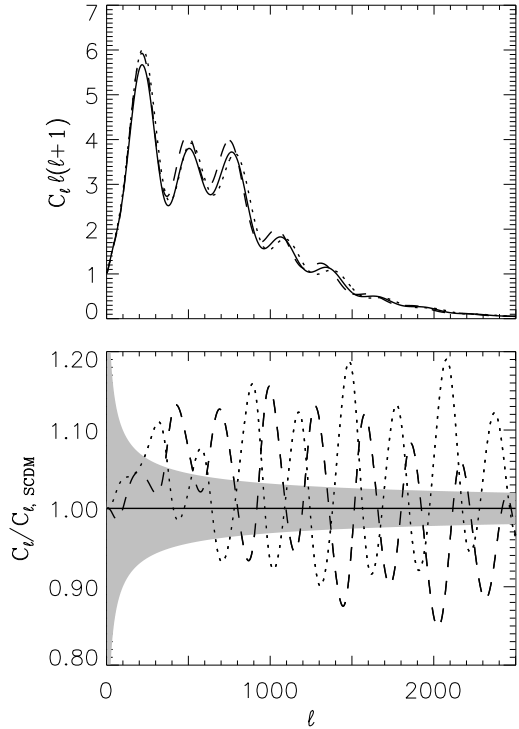}\hss}
\caption{{\it Top:} CMBR fluctuation spectrum for SCDM with $h=0.5$,
  $\Omega_M=1$, $\Omega_B=0.05$, and $N_{\rm eff}=3$ (solid
  line)~\protect\cite{Hannestad99}.  The dotted line is for $N_{\rm
  eff}=4$, and the dashed line when two of these four neutrinos have
  equal masses corresponding together to $\Omega_{\rm HDM}=0.2$
  ($\Omega_{\rm CDM}=0.75$).  {\it Bottom:}~Relative difference of
  these nonstandard models to SCDM.  The shaded band represents the
  cosmic variance.  (Spectra calculated with the
  CMBFAST~\protect\cite{CMBFAST} package.)}
\label{fig:cmbr}
\end{figure}

CMBR sky maps are characterized by their fluctuation spectrum
$C_\ell=\langle a^{}_{\ell m} a^*_{\ell m}\rangle$ where $a_{\ell m}$
are the coefficients of a spherical-harmonic expansion.
Figure~\ref{fig:cmbr} (solid line) shows $C_\ell$ for standard cold
dark matter (SCDM) with $N_{\rm eff}=3$ for the effective number of
neutrino degrees of freedom.  Sterile neutrinos increase the radiation
content and thus modify this pattern in a characteristic way
illustrated by the dotted line, which corresponds to $N_{\rm eff}=4$.

While this shift appears small, the lower panel of Fig.~\ref{fig:cmbr}
shows that for $\ell\agt 200$ it is large on the scale of the expected
measurement precision. It is fundamentally limited by the ``cosmic
variance'' $\Delta C_\ell/C_\ell=\sqrt{2/(2\ell+1)}$, i.e.\ by the
fact that at our given location in the universe we can measure only
$2\ell+1$ numbers $a_{\ell m}$ to obtain the expectation value
$\langle a^{}_{\ell m} a^*_{\ell m}\rangle$.  The actual sensitivity
will be worse, but the cosmic variance gives us an optimistic idea of
what one may hope to achieve.
The true sensitivity to $\Delta N_{\rm eff}$ is further limited by our
lack of knowledge of several other cosmological parameters.  Even then
it is safe to assume that we are sensitive to $|\Delta N_{\rm
eff}|\alt0.3$, and much better with prior knowledge of other
parameters~\cite{Jungman96b}. Thus it appears that the CMBR is a more
powerful tool to measure $N_{\rm eff}$ than the standard BBN argument,
although a more pessimistic assessment was put forth in a more recent
analysis~\cite{Hu98c}.

If LSND is right, some of the neutrinos have eV masses which imprint
themselves on the CMBR fluctuation spectrum~\cite{Ma95,Dodelson96}.
For example, if the atmospheric neutrino anomaly is due to
$\nu_\mu$-$\nu_s$-oscillations, we will have approximately $N_{\rm
eff}=4$, and two of these states will have an eV-range mass. The CMBR
imprint of this scenario is illustrated with the dashed curve in
Fig.~\ref{fig:cmbr} where $\Omega_\nu=0.2$.  With
$\Omega_{2\nu}h^2=2m_\nu/93~{\rm eV}$ and taking $h=0.5$ this implies
$m_\nu\approx2.4~{\rm eV}$, well within the range suggested by~LSND.

The range of $\Delta N_{\rm eff}$ and the HDM fraction that can be
determined by the future CMBR sky maps, together with large-scale
galaxy surveys, cannot be foretold with certainty, but surely these
cosmological precision observables are significantly affected by the
currently debated neutrino mass and mixing patterns. Cosmology may be
our best bet to pin down the overall neutrino mass scale which is left
undetermined by oscillation experiments.


\section{Supernova Physics}
\label{sec:supernova}

\subsection{Kinematical Mass Limits}

When SN~1987A exploded on 23 February 1987 in the Large Magellanic
Cloud at a distance of about 50~kpc (165,000~lyr), it produced the
third case of a measured neutrino signal from an astrophysical source
after the Sun and the Earth's atmosphere.  Therefore, we turn to the
role of masses and mixings for SN neutrinos in general, and for the
SN~1987A burst in particular.

A type~II SN explosion~\cite{Brown82,Bethe90,Petschek90} marks
the end of the life of a massive star ($M\agt8\,M_\odot$) which has
developed a degenerate iron core, surrounded by several burning
shells. As the core reaches its Chandrasekhar limit of 1--$2\,M_\odot$
(solar masses) it becomes unstable and collapses down to nuclear
density ($3\times10^{14}~{\rm g~cm^{-3}}$) where the equation of state
stiffens and the implosion is halted.  At this point a 
shock wave forms which ejects the mantle of the
progenitor star---the SN explosion is the reversed core implosion.

At about nuclear density and a temperature of several 10~MeV 
the newly formed neutron star is opaque to neutrinos which are
thus emitted from a shell at about unit optical depth,
the ``neutrino sphere,'' crudely with a thermal spectrum. One expects
that the total binding energy~\cite{Sato87,Janka89}
\begin{equation}\label{eq:bindingenergy}
E_{\rm b}=\hbox{1.5--4.5}\times10^{53}~{\rm erg}
\end{equation}
is roughly equipartioned between all (anti)neutrino flavor degrees of
freedom and that it is emitted within several seconds. This picture
agrees well with the SN~1987A observations in the
Kamiokande~\cite{Hirata88} and IMB~\cite{Bratton88} water Cherenkov
detectors and the Baksan Scintillator Telescope~\cite{Alexeyev87}
which were all primarily sensitive to the positrons from the
$\bar\nu_e+p\to n+e^+$ capture reaction.

A neutrino mass can manifest itself by a time-of-flight
dispersion of the SN burst~\cite{Zatsepin68}.  The 
neutrino arrival time
from a distance $D$ is delayed by
\begin{equation}\label{eq:sndelay}
\Delta t=2.57~{\rm s}\,
\left(\frac{D}{50~{\rm kpc}}\right)\,
\left(\frac{10~{\rm MeV}}{E_\nu}\right)^2\,
\left(\frac{m_\nu}{10~{\rm eV}}\right)^2.
\end{equation}
As the 
$\bar\nu_e$'s from SN~1987A 
were registered within a few seconds and had
energies in the 10~MeV range, the $m_{\nu_e}$ limit is 
around 10~eV.  Detailed analyses reveal that the pulse duration
is consistently explained by the SN cooling time 
and that $m_{\nu_e}\alt 20~{\rm eV}$ is implied at something
like 95\% CL~\cite{Loredo89,Kernan95}.

The high-statistics observation of a future galactic SN with a large
detector like SuperKamiokande allows one to improve the
$m_{\nu_e}$-sensitivity to about $3~{\rm eV}$ because one can use the
fast rise-time of the signal as a dispersion measure rather than the
overall burst duration itself~\cite{Totani98}.  On the other hand, the
neutral-current signal in a large water Cherenkov detector like
SuperKamiokande or SNO provides a direct handle on $m_{\nu_\mu}$ and
$m_{\nu_\tau}$ of no better than
30~eV~\cite{Seckel91,Krauss92,Fiorentini97a,Beacom98}.  Even with a
future neutral-current detector like OMNIS it is not realistically
possible to probe $m_{\nu_\mu}$ and $m_{\nu_\tau}$ down to a few
eV~\cite{Cline94,Smith97}.

\subsection{SN~1987A and Flavor Oscillations}

While the SN~1987A limit on $m_{\nu_e}$ is not truly interesting for
the current debate, the event energies bear on the large-angle
solutions of the solar neutrino problem, and especially on the vacuum
solution.  In typical numerical simulations one finds for the average
energies for the different flavors~\cite{Janka93}
\begin{equation}\label{eq:energies}
\langle E_{\nu}\rangle=\cases{10{-}12\,{\rm MeV}&for $\nu_e$,\cr
14{-}17\,{\rm MeV}&for $\bar\nu_e$,\cr
24{-}27\,{\rm MeV}&for $\nu_{\mu,\tau}$ and
$\bar\nu_{\mu,\tau}$,}
\end{equation}
so that $\langle E_{\nu_e}\rangle:\langle E_{\bar\nu_e}\rangle:
\langle E_{\rm others}\rangle\approx \frac{2}{3}:1:\frac{5}{3}$.
Large mixing angle oscillations between $\bar\nu_e$ and $\bar\nu_\mu$
would partially swap their fluxes and thus ``stiffen'' the $\bar\nu_e$
spectrum observable at
Earth~\cite{Kernan95,Wolfenstein87,Lagage87,Smirnov94,Jegerlehner96}.
(We take $\bar\nu_\mu$ to stand for either $\bar\nu_\mu$
or~$\bar\nu_\tau$.)  Therefore, some of the SN~1987A events would have
been oscillated $\bar\nu_\mu$'s which should have been correspondingly
more energetic.

\begin{figure}[ht]
\epsfxsize=7cm \hbox to\hsize{\hss\epsfbox{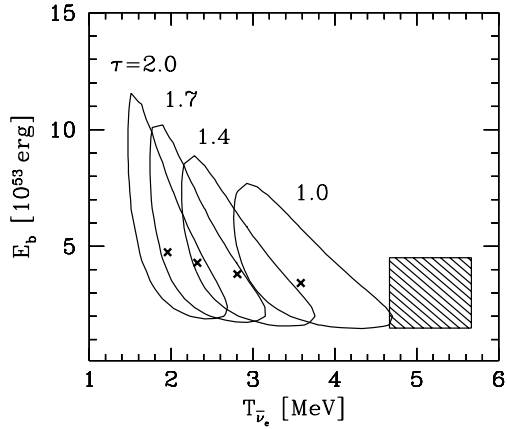}\hss}
\caption{Best-fit values for the spectral $\bar\nu_e$ temperature
$T_{\bar\nu_e}$ and the neutron-star binding energy $E_{\rm b}$, as
well as contours of constant likelihood corresponding to 95\%
confidence regions~\protect\cite{Jegerlehner96}.  They are based on a
joint analysis between the Kamiokande and IMB data, assuming maximum
mixing and the indicated values for
$\tau=T_{\bar\nu_\mu}/T_{\bar\nu_e}$, where $\tau=1$ corresponds to no
oscillations.  The hatched region represents the predictions of
Eqs.~(\protect\ref{eq:bindingenergy}) and~(\protect\ref{eq:energies}).
\label{fig:sn}}
\end{figure}

A maximum-likelihood analysis of the $\bar\nu_e$ spectral temperature
and the neutron-star binding energy inferred from the
Kamiokande~\cite{Hirata88} and IMB~\cite{Bratton88} data
(Fig.~\ref{fig:sn}) reveals that even in the no-oscillation case there
is only marginal overlap with the theoretical expectation of
Eq.~(\protect\ref{eq:energies}).  The observed neutrinos were softer
than predicted, especially at Kamiokande. Including a spectral swap
exacerbates this problem in that the energies should have been even
higher. In Fig.~\ref{fig:sn} we show 95\% likelihood contours for the
infered $\bar\nu_e$ spectral temperature $T_{\bar\nu_e}=\langle
E_{\bar\nu_e}\rangle/3$ and the neutron-star binding energy $E_{\rm
b}$ for maximum $\bar\nu_e$-$\bar\nu_\mu$-mixing and for several
values of $\tau=T_{\bar\nu_\mu}/T_{\bar\nu_e}$.  Even for moderate
spectral differences a maximum mixing between $\bar\nu_e$ and the
other flavors causes a conflict with the SN~1987A
data~\cite{Smirnov94,Jegerlehner96}.

It may be premature to exclude the solar vacuum solution on these
grounds as the spectral differences may have been overestimated.  They
arise because of flavor-dependent opacities. The electron-flavored
neutrinos are trapped by $\nu_e n\to p e^-$ and $\bar\nu_e p\to n
e^+$.  The other flavors interact by neutral-current collisions which
have smaller cross sections so that these particles emerge from deeper
and hotter layers.  They escape from their ``transport sphere'' where
collisions are no longer effective, but most critical for their
spectrum is the ``energy sphere'' where they last exchanged energy
with the medium~\cite{Janka95}.  Electron scattering $\nu e^-\to
e^-\nu$ was taken to dominante for energy-exchange and $e^+e^-\to
\nu\bar\nu$ for pair production. However, the dominant pair-process is
nucleonic bremsstrahlung~\cite{Suzuki91,Hannestad98} $NN\to
NN\nu\bar\nu$, the dominant energy-exchange processes are recoils and
inelasticities in $\nu N\to N\nu$
scattering~\cite{Janka96,Hannestad98}.  Including these effects
clearly makes the $\bar\nu_\mu$ spectrum more similar to
$\bar\nu_e$. A preliminary estimate suggests that the remaining
spectral differences may be small enough to avoid a conflict between
SN~1987A and the solar vacuum solution~\cite{Hannestad98}.  Since
neutrino oscillations can be crucial for the interpretation of the
signal from a future galactic SN~\cite{Qian94,Choubey98,Fuller98}, one
should indeed spend more effort at understanding details of the
spectra formation process~\cite{Hardy99}.

An interesting case which does not depend on the spectral
differences is the ``prompt $\nu_e$ burst,'' originating from the
deleptonization of the outer core layers at about 100~ms after bounce
when the shock wave breaks through the edge of the collapsed
core.  This ``deleptonization burst'' propagates through the mantle
and envelope of the progenitor star so that resonant oscillations take
place for a large range of mixing parameters between $\nu_e$ and some
other flavor, notably for some of those values where the MSW effect
operates in the Sun~\cite{Mikheyev86,Notzold87,Rosen88}.  In a 
Cherenkov detector one can see this burst by $\nu_e$-$e$-scattering
which is forward peaked, but one would have expected only a fraction
of an event from SN~1987A.  The first event in Kamiokande may be
attributed to this signal, but this interpretation is statistically
insignificant. The experimental signal of the prompt $\nu_e$ burst
from a future galactic SN is closely intertwined with the mixing
parameters which solve the solar neutrino problem.

\subsection{Flavor Oscillations and Supernova Physics}

Flavor oscillations can have interesting ramifications for SN physics
itself, independently of neutrino flux measurements at Earth. As
galactic SNe are rare (one every few decades or even less) it is not
guaranteed that we will observe neutrinos from another SN anytime
soon.  Therefore, it is even more important to use the SN phenomenon
itself as a laboratory for neutrino physics.

\begin{figure}[b]
\hbox to\hsize{\hss\epsfxsize=7cm\epsfbox{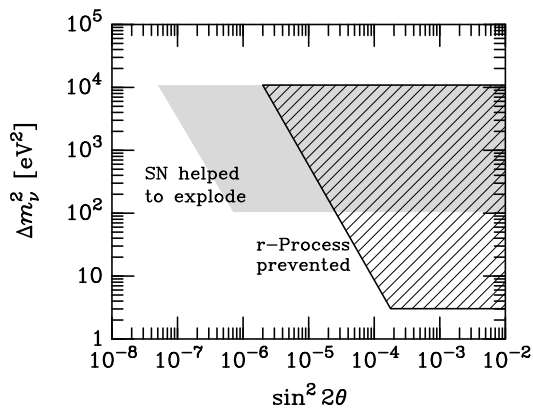}\hss}
\caption{Mixing parameters between $\nu_e$ and $\nu_\mu$ or $\nu_\tau$
 where a spectral swap would help explode
 supernovae~\protect\cite{Fuller92} and where it would prevent
 r-process nucleosynthesis~\protect\cite{Qian93,Qian95,Sigl95}.
\label{fig:snosci}}
\end{figure}

For example, flavor oscillations can help with the
explosion~\cite{Fuller92}.  The standard scenario of a type~II SN
explosion has it that a shock wave forms near the edge of the core
when its collapse halts at nuclear density and that this shock wave
ejects the mantle of the progenitor star. However, in typical
numerical calculations the shock wave stalls so that this ``prompt
explosion'' scenario does not seem to work. In the ``delayed
explosion'' picture the shock wave is revived by neutrino heating,
perhaps in conjunction with convection, but even then it appears
difficult to obtain a successful or sufficiently energetic explosion.
The efficiency of neutrino heating can increase by resonant flavor
oscillations which swap the $\nu_e$ flux with, say, the $\nu_\tau$
one.  Therefore, what passes through the shock wave as a $\nu_e$ was
born as a $\nu_\tau$ at the proto neutron star surface. It has on
average higher energies and thus is more effective at transfering
energy. In Fig.~\ref{fig:snosci} the shaded range of mixing parameters
is where SNe are helped to explode, assuming a ``normal'' neutrino
mass spectrum with $m_{\nu_e}<m_{\nu_\tau}$.  Below the shaded region
the resonant oscillations take place beyond the shock wave and thus do
not affect the explosion.

A few seconds after core bounce the shock wave has long taken off,
leaving behind a relatively dilute ``hot bubble'' above the
neutron-star surface. This region is one suspected site for the
r-process heavy-element synthesis, which requires a neutron-rich
environment~\cite{Woosley92,Witti94,Hoffman96,Meyer97,%
Meyer98}.  The neutron-to-proton ratio, which is governed by the beta
reactions $\nu_e+n\to p+e^-$ and $\bar\nu_e+p\to n+e^+$, is shifted to
a neutron-rich phase if $\langle E_{\nu_e}\rangle<\langle
E_{\bar\nu_e}\rangle$ as for standard neutrino spectra.  Resonant
oscillations can again swap the $\nu_e$ flux with another one,
inverting this hierarchy of energies.  In the hatched range of mixing
parameters shown in Fig.~\ref{fig:snosci} the r-process would be
disturbed~\cite{Qian93,Qian95,Sigl95}, in conflict with the
upper range of LSND-inspired mass differences.
On the other hand, oscillations $\nu_e\to\nu_s$ into a sterile
neutrino could actually help the r-process by depleting
the neutron-stealing $\nu_e$ flux~\cite{Nunokawa97a,Caldwell98}.

\subsection{Pulsar Kicks by Oscillations?}

Radio pulsars often move with
velocities~\cite{Lyne94,Lorimer97,Cordes98} of several $100~{\rm
km~s^{-1}}$, a phenomenon yet to be explained.  The acceleration
probably takes place in the context of their formation in a
core-collapse SN, i.e.\ they likely receive a kick at birth. One
explanation appeals to a ``neutrino rocket'' because the
momentum carried by the neutrino burst is so large that an emission
anisotropy as small as 1\% suffices to account for a recoil of about
$300~{\rm km~s^{-1}}$.  However, even such a small anisotropy is
difficult to explain.

Pulsars tend to have strong magnetic fields which may well be
suspected to cause the asymmetry.  The neutrino refractive index
depends on the direction of the neutrino momentum relative to ${\bf
B}$. For suitable conditions, resonant neutrino oscillations occur
between the neutrinospheres of $\nu_e$ and $\nu_\tau$, deforming the
effective $\nu_\tau$~sphere.  The $\nu_\tau$'s would thus emerge from
regions of varying effective temperature and thus, it was argued,
would be emitted anisotropically~\cite{Kusenko96}.  This argument was
then taken up in several papers with modified neutrino oscillation
scenarios~\cite{Kusenko97,Akhmedov97a,Grasso98,Horvat98}.

Unfortunately, this intruiging idea does not work for plausible
magnetic field strengths~\cite{Janka99}.  The oscillations take place
in the ``atmosphere'' of the neutron star, while the neutrino flux is
fixed much deeper inside. The atmosphere adjusts itself to transport
the neutrino flux, not the other way round.  Neutrino oscillations in
the atmosphere leave the overall flux unchanged except for a
higher-order backreaction effect which obtains because of the
anisotropically modified atmospheric structure.  It may still be that
a neutrino rocket effect is responsible for the pulsar kicks, but the
cause for the anisotropy remains unclear and if it is related to
nonstandard neutrino properties.

\subsection{Neutrino Mass Limit from Neutron-Star Stability?}

In a thought-provoking paper~\cite{Fischbach96} it was recently
claimed that neutron stars provided a {\it lower\/} neutrino mass
limit of $m_\nu\agt0.4~{\rm eV}$.  Two-neutrino exchange between
fermions gives rise to a long-range force.  A neutrino may also pass
around several fermions, so to speak, producing a much smaller
potential.  This multibody neutrino exchange, it was argued, would be
a huge effect in neutron stars because combinatorial factors among
many neutrons win out against the smallness of the potential for a
given set of them.  One way out is to suppress the long-range nature
of neutrino exchange by a nonzero $m_\nu$.

This idea triggered a series of papers where it was shown that a
proper resummation of a seemingly divergent series of terms leads to a
well-behaved and small ``neutron-star self-energy,'' invalidating the
claim of a lower neutrino mass
limit~\cite{Kiers98,Abada96,Abada98,Arafune98}.  As naively expected,
there is no mysterious long-range force from neutrino exchange, but
these papers are still interesting reading for anyone interested in
questions of neutrino physics in media.


\section{Neutrino Astronomy}
\label{sec:neutrinoastronomy}

\subsection{Neutrino Telescopes}

For twenty years after the first observation of solar neutrinos at the
Homestake detector, neutrino astronomy remained a one-experiment
field. The SN~1987A neutrino observations mark a turning point---the
number of experiments and observatories has multiplied since about
that time, with more than a dozen previous, operating or projected
neutrino detectors measuring solar and atmospheric neutrinos or
searching for a new SN burst. The neutrino sky at low energies is
dominated by these sources with a solar $\nu_e$ flux of around
$6.6\times10^{10}~{\rm cm^{-2}~s^{-1}}$ in the MeV range and that from
a SN at a distance of 10~kpc of around $3\times10^{12}~{\rm
  cm^{-2}~s^{-1}}$ in the 10--100~MeV range during the burst of a few
seconds.  At around 1~GeV the atmospheric neutrino flux for all
flavors together and integrated over all angles is $dN_\nu/d\ln
E_\nu\approx0.7~{\rm cm^{-2}~s^{-1}}$, dropping with energy
approximately as $E_\nu^{-2}$.

A new development is the emergence of huge neutrino telescopes with
the goal of observing astrophysical sources of neutrinos with energies
in the TeV range and beyond~\cite{Gaisser95,Halzen97a,Halzen98a}.  The
existence of cosmic rays with energies reaching beyond $10^{20}~{\rm
eV}$ proves that they must have been accelerated somewhere, but the
nature of the accelerators remains mysterious. Protons are deflected
in the micro-Gauss galactic magnetic field so that the cosmic rays
hitting the Earth do not point back to their sources, a problem not
shared by neutrinos. High-energy neutrinos are expected from ``cosmic
beam dumps'' whenever the protons interact with matter or even photons
to produce pions---the Earth's atmosphere as a neutrino source is the
simplest case in point.

\begin{figure}
\hbox to\hsize{\hss\epsfxsize=11.8cm\epsfbox{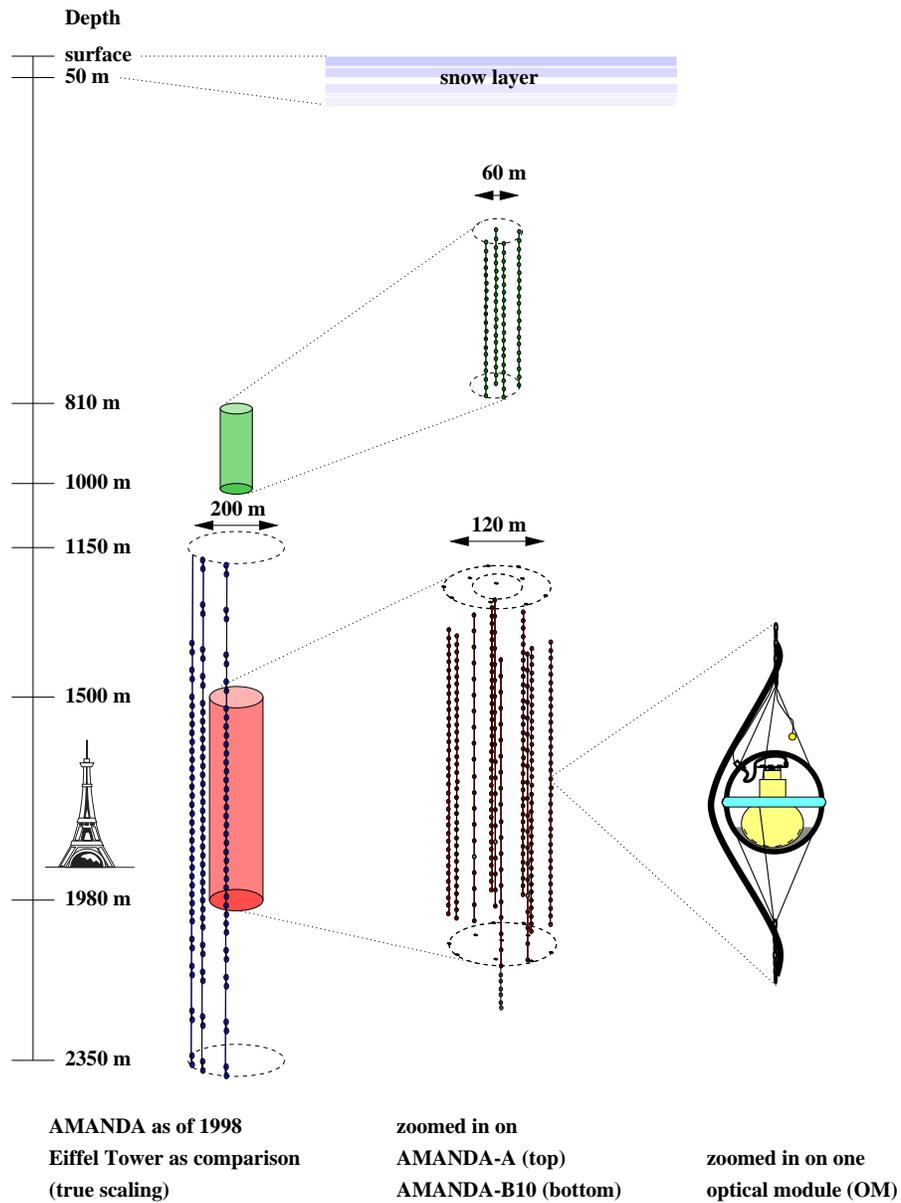}\hss}
\caption{Schematic view of the AMANDA
  South Pole high-energy neutrino telescope~\protect\cite{Amanda98}.
  (Figure reproduced with permission of F.~Halzen.)
\label{fig:amanda}}
\end{figure}

Estimates of the expected neutrino fluxes vary, but certainly one
needs detectors far exceeding the size of SuperKamiokande. For a
useful neutrino Cherenkov telescope one probably needs a
cubic-kilometer of water or ice instrumented with photomultipliers
which can be placed on a grid with a typical spacing of order 30~m.
There are now several such utopian-sounding projects on their way.  A
small but functioning instrument has been deployed in Lake
Baikal~\cite{Baikal97} but probably it will not grow to the ${\rm
km^3}$ scale. Two Mediterranean projects, NESTOR~\cite{Nestor99} and
ANTARES~\cite{Antares98}, are in the R\&D and feasibility-study phase.
At present the most advanced detector with a realistic ${\rm km}^3$
perspective is AMANDA~\cite{Amanda98} at the South Pole
(Fig.~\ref{fig:amanda}).  The antarctic ice is used both as a
Cherenkov medium and as a mechanical support structure for strings of
photomulipliers which are frozen into 2~km deep holes.

The main focus of these exciting projects is neutrino astronomy, i.e.\
to study the sky in a new form of radiation and to learn about the
nature of the astrophyscial sources. However, high-energy neutrino
astronomy has several important ramifications of direct
particle-physics interest.

\subsection{Search for Particle Dark Matter}

First, one may search for dark matter in the form of weakly
interacting massive particles (WIMPs), especially in the guise of the
supersymmetric neutralinos. The case for these particles has become
stronger as massive neutrinos no longer seem tenable as a main
dark-matter constituent. Galactic WIMPs are accreted by the Sun or
Earth where they annihilate with each other, leading to a secondary
GeV--TeV neutrino flux. Depending on details of the assumed
supersymmetric model, this ``indirect'' method to search for particle
dark matter is competitive with the direct laboratory
experiments~\cite{Jungman96,Bergstrom97}.

\subsection{Tau-Neutrinos from Astrophysical Sources}

Neutrinos produced in cosmic beam dumps should have the same flavor
content as those produced in the atmosphere. If atmospheric neutrinos
indeed oscillate, so do the ones from high-energy astrophysical
sources. If the $\nu_\mu\to\nu_\tau$ oscillation channel is what
explains the atmospheric anomaly, then the astrophysical beam dumps
produce a flux which includes high-energy $\nu_\tau$'s.  

One signature in a Cherenkov detector are so-called double-bang
events~\cite{Learned95} which consist of a big hadronic shower from
the initial $\nu_\tau$ interaction, a muon-like $\tau$-track, and then
a second big particle cascade when the $\tau$ decays. This could be
100~m downstream from the first interaction if the primary energy was
in the PeV ($10^{15}~{\rm eV}$) range as expected from active galactic
nuclei (AGNs) as neutrino sources~\cite{Halzen97b}.  However, such
signatures may be difficult to detect in a first-generation telescope
like AMANDA.

The Earth is opaque to neutrinos with energies above something like
100~TeV, but $\nu_\tau$'s can still make it to the detector from
below~\cite{Halzen98b}.  The main idea is that a $\tau$ produced in a
charged-current interaction of the primary $\nu_\tau$ decays back into
a $\nu_\tau$ before losing much energy, thereby piling up $\nu_\tau$'s
at energies around 100~TeV. Moreover, this effect would manifest
itself by a flat zenith-angle dependence of source intensity at the
highest energies~\cite{Halzen98b}.  

The atmospheric neutrino anomaly has rather immediate consequences for
high-energy neutrino astronomy!

\subsection{Neutrino Masses}

Besides AGNs, gamma-ray bursts are one of the favored suspects for
producing the highest energy cosmic rays and for producing high-energy
neutrinos~\cite{Waxman97}. Their pulsed nature allows one to search
for neutrino masses by time-of-flight dispersion in analogy to the
SN~1987A mass limit.  Since typical gamma-ray bursts are at
cosmological distances of order 1000~Mpc, one gains enormously in
Eq.~(\ref{eq:sndelay}) relative to SN~1987A, but of course the final
mass sensitivity depends on the time-structure (perhaps as short as
milliseconds) and the observed neutrino energies.

If neutrinos with energies as high as $10^{22}~{\rm eV}$ are copiously
produced in astrophysical sources, and if eV-mass neutrinos exist as a
hot-dark matter component and are locally clustered, then high-energy
particle cascades would be initiated which could produce, as secondary
products, the highest-energy observed cosmic rays which have energies
beyond $10^{20}~{\rm eV}$~\cite{Weiler97,Fargion97,Yoshida98}.  The
universe is opaque for protons above $4\times10^{19}~{\rm eV}$, the
Greisen-Zatsepin-Kuzmin cutoff, due to photo-pion production on the
cosmic microwave radiation.  Therefore, the highest-energy cosmic
rays, if they are protons, must have a local source, but the observed
events do not point toward any plausible structure which might serve
as such.  Neutrinos thus offer one of many speculative explanations
for the puzzle of the highest-energy cosmic rays.


\section{Neutrino Electromagnetic Properties}
\label{sec:electromagneticproperties}

\subsection{Form Factors}

A survey of neutrino astrophysics would be incomplete without a
discussion of neutrino electromagnetic properties which could have
several important astrophysical consequences.  The most general
neutrino interaction with the electromagnetic field
is~\cite{Mohapatra91,Winter91}
\begin{equation}
{\cal L}_{\rm int}=-F_1\bar\psi\gamma_\mu\psi A^\mu
-G_1\bar\psi\gamma_\mu\gamma_5\psi\partial_\mu F^{\mu\nu}
-{\textstyle\frac{1}{2}}
\bar\psi\sigma_{\mu\nu}(F_2+G_2\gamma_5)\psi F^{\mu\nu},
\end{equation}
where $\psi$ is the neutrino field, $A^\mu$ the electromagnetic vector
potential, and $F^{\mu\nu}$ the field-strength tensor.  The form
factors are functions of $Q^2$ with $Q$ the energy-momentum transfer.
In the $Q^2\to0$ limit $F_1$ is a charge, $G_1$ an anapole moment,
$F_2$ a magnetic, and $G_2$ an electric dipole moment.

Charge neutrality implies $F_1(0)=0$.  What remains is a charge radius
which, like the anapole moment, vanishes in the $Q^2\to0$ limit.
Therefore, it provides for a contact interaction and as such a
correction to processes with $Z^0$
exchange~\cite{Degrassi89,Gongora92}.  As astrophysics provides no
precision test for the effective strength of neutral-current
interactions, these form factors are best probed in laboratory
experiments~\cite{Salati94}.

Therefore, the only astrophysically interesting possibility are
magnetic and electric dipole and transition moments.  If the standard
model is extended to include neutrino Dirac masses, the magnetic
dipole moment is $\mu_\nu=3.20\times10^{-19}\,\mu_{\rm B}\,m_\nu/{\rm
eV}$ where $\mu_{\rm B}=e/2m_e$ is the Bohr
magneton~\cite{Mohapatra91,Winter91}.  An electric dipole moment
$\epsilon_\nu$ violates CP, and both are forbidden for Majorana
neutrinos.  Flavor mixing implies electric and magnetic transition
moments for both Dirac and Majorana neutrinos, but they are even
smaller due to a GIM cancelation.  Neutrino electromagnetic form
factors which are large enough to be of experimental or astrophysical
interest require a more radical extension of the standard model, for
example the existence of right-handed currents.

\subsection{Astrophysical Limits}

\begin{figure}
\hbox to\hsize{\hss\epsfxsize=7cm\epsfbox{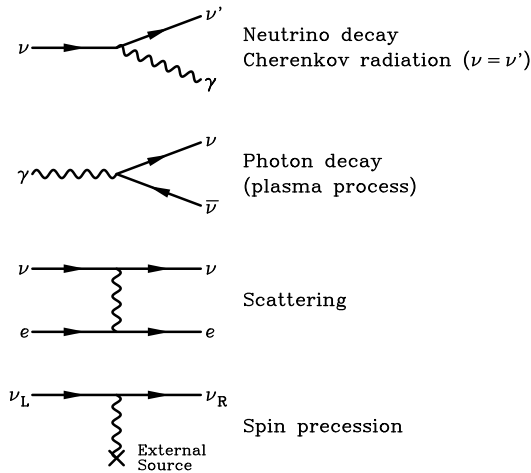}\hss}
\caption{Processes with neutrino electromagnetic
dipole or transition moments.\label{fig:processes}}
\end{figure}

Assuming that neutrinos have nonstandard electric or magnetic dipole
or transition moments, how large can they be?  Astrophysics, not
laboratory experiments, provides the most restrictive limits.  Dipole
or transition moments allow for several interesting processes
(Fig.~\ref{fig:processes}).  For the purpose of deriving limits, the
most important case is $\gamma\to\nu\bar\nu$ which is kinematically
possible in a plasma because the photon acquires a dispersion relation
which roughly amounts to an effective mass.  Even without anomalous
couplings, the plasmon decay proceeds because the charged particles of
the medium provide an effective neutrino-photon
interaction~\cite{Adams63,Zaidi65,Haft94}.  Put another way, even
standard neutrinos have nonvanishing electromagnetic form factors in a
medium~\cite{DOlivo89,Altherr94}.

The standard plasma process dominates the neutrino production in white
dwarfs or the degenerate helium core of globular-cluster red
giants. The presence of a direct neutrino-photon coupling by a dipole
or transition moment enhances the neutrino losses, delaying the
ignition of helium.  Observations of globular-cluster stars thus
reveal a
limit~\cite{Raffelt90,Raffelt92,Castellani93,Catelan96,Raffelt96}
\begin{equation}\label{eq:dipolelimit}
\mu_\nu\alt3\times10^{-12}\,\mu_{\rm B},
\end{equation}
applicable to magnetic and electric dipole and transition moments for
Dirac and Majorana neutrinos.  Of course, the final-state neutrinos
must be lighter than the photon plasma mass of around 10~keV for the
relevant conditions. A slightly weaker bound obtains from the
white-dwarf luminosity function~\cite{Blinnikov94}.  Right-handed
(sterile) states are produced in electromagnetic spin-flip collisions
if neutrinos have Dirac dipole or transition moments.  The duration of
the SN~1987A neutrino signal precludes excessive cooling by sterile
states, yielding a limit on $\mu_\nu({\rm Dirac})$ which is
numerically equivalent to
Eq.~(\ref{eq:dipolelimit})~\cite{Barbieri88a,Ayala98}.

The corresponding laboratory limits are much weaker~\cite{Caso98}.
The most restrictive bound is $\mu_{\nu_e}<1.8\times10^{-10}\,\mu_{\rm
B}$ at 90\% CL from a measurement of the $\bar\nu_e$-$e$-scattering
cross section involving a reactor source. A significant improvement
should become possible with the MUNU experiment~\cite{Broggini99}, but
it is unlikely that the globular-cluster limit can be reached anytime
soon.

\begin{figure}[b]
\hbox to\hsize{\hss\epsfxsize=7cm\epsfbox{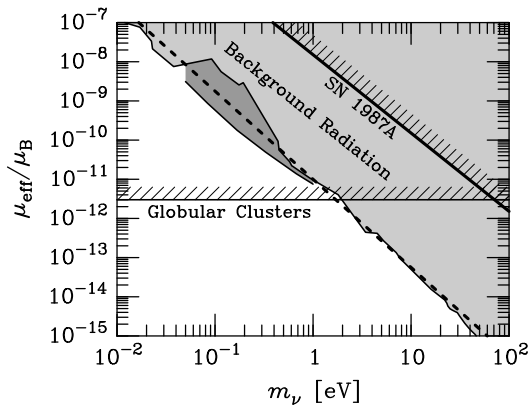}\hss}
\caption{Astrophysical limits on neutrino dipole moments. The
light-shaded~\protect\cite{Ressell90} and
dark-shaded~\protect\cite{Biller98,Raffelt98} exclusion range is from
the absence of excessive cosmic diffuse background photons.  The
dashed line represents the approximation formula in
Eq.~(\protect\ref{eq:munulimits}), bottom line. 
\label{fig:munu}}
\end{figure}

A neutrino mass eigenstate $\nu_i$ may decay to another one $\nu_j$ by
the emission of a photon, where the only contributing form factors are
the magnetic and electric transition moments. The inverse radiative
lifetime is found to be~\cite{Mohapatra91,Winter91}
\begin{eqnarray}\label{eq:radiativedecay}
\tau_\gamma^{-1}&=&\frac{|\mu_{ij}|^2+|\epsilon_{ij}|^2}{8\pi}
\left(\frac{m_i^2-m_j^2}{m_i}\right)^3\nonumber\\
&=&5.308~{\rm s}^{-1}
\left(\frac{\mu_{\rm eff}}{\mu_{\rm B}}\right)^2
\left(\frac{m_i^2-m_j^2}{m_i^2}\right)^3
\left(\frac{m_i}{{\rm eV}}\right)^3,
\end{eqnarray}
where $\mu_{ij}$ and $\epsilon_{ij}$ are the transition moments while
$|\mu_{\rm eff}|^2\equiv|\mu_{ij}|^2+|\epsilon_{ij}|^2$.  Radiative
neutrino decays have been constrained from the absence of decay
photons of reactor $\bar\nu_e$ fluxes~\cite{Oberauer87}, the solar
$\nu_e$ flux~\cite{Raffelt85}, and the SN~1987A neutrino
burst~\cite{Feilitzsch88,Chupp89,Kolb89,Bludman92,Oberauer93}.  For
$m_\nu\equiv m_i\gg m_j$ these limits can be expressed as
\begin{equation}\label{eq:munulimits}
\frac{\mu_{\rm eff}}{\mu_{\rm B}}\;\alt\;
\cases{\hbox to1.8cm{$0.9{\times}10^{-1}$\hfil}({\rm eV}/m_\nu)^2
&Reactor ($\bar\nu_e$),\cr
\hbox to1.8cm{$0.5{\times}10^{-5}$\hfil}({\rm eV}/m_\nu)^2
&Sun ($\nu_e$),\cr
\hbox to1.8cm{$1.5{\times}10^{-8}$\hfil}({\rm eV}/m_\nu)^2
&SN~1987A (all flavors),\cr
\hbox to1.8cm{$1.0{\times}10^{-11}$\hfil}({\rm eV}/m_\nu)^{9/4}
&Cosmic background (all flavors).\cr}
\end{equation}
In this form the SN~1987A limit applies for $m_\nu\alt 40~{\rm eV}$.
The decay of cosmic background neutrinos would contribute to the
diffuse photon backgrounds, excluding the shaded areas in
Fig.~\ref{fig:munu}. They are approximately delineated by the dashed
line, corresponding to the analytic expression in
Eq.~(\ref{eq:munulimits}). More restrictive limits obtain for certain
masses above 3~eV from the absence of emission features from several
galaxy clusters~\cite{Henry81,Davidsen91,Bershady91}.

For low-mass neutrinos the $m_\nu^3$ phase-space factor in
Eq.~(\ref{eq:radiativedecay}) is so punishing that the
globular-cluster limit is the most restrictive one for $m_\nu$ below a
few eV, i.e.\ in the mass range which today appears favored from
neutrino oscillation experiments.  Turning this around, if neutrino
mass differences are indeed as small as currently believed, the
globular-cluster limit implies that radiative neutrino decays do not
have observable consequences.

\subsection{Spin and Spin-Flavor Precession}

Neutrinos with magnetic or electric dipole moments spin-precess in
external magnetic fields~\cite{Fujikawa80,Okun86a}, an effect which
may have a number of astrophysical consequences for $\mu_\nu$-values
below the globular-cluster limit of Eq.~(\ref{eq:dipolelimit}).  For
example, solar neutrinos can precess into sterile and thus
undetectable states in the Sun's magnetic
field~\cite{Werntz70,Cisneros71,Voloshin86a}.  The same for SN
neutrinos in the galactic magnetic field where an important effect
obtains for $\mu_\nu\agt10^{-12}\,\mu_{\rm B}$. Moreover, the
high-energy sterile states emitted by spin-flip collisions from the
inner SN core could precess back into active ones and cause events
with anomalously high energies in SN neutrino detectors, an effect
which probably requires $\mu_\nu({\rm Dirac})\alt10^{-12}\,\mu_{\rm
B}$ from the SN~1987A signal~\cite{Barbieri88a,Notzold88}.  For the
same $\mu_\nu$-range one may expect an anomalous rate of energy
transfer to the shock wave in a SN, helping with the explosion
\cite{Dar87c,Nussinov87a,Goldman88,Voloshin88,Okun88a,Blinnikov88}.

The refractive energy shift in a medium for active neutrinos relative
to sterile ones creates a barrier to spin
precessions~\cite{Voloshin86b}.  The neutrino mass difference has the
same effect if the precession is between different flavors through a
transition moment~\cite{Schechter81}.  Combining the effects one
arrives at spin-flavor precession in a medium.  The mass difference
and the refractive term can cancel, leading to resonant oscillations
in the spirit of the MSW
effect~\cite{Akhmedov88a,Akhmedov88b,Barbieri88b,Lim88}.

Large magnetic fields exist in SN cores so that spin-flavor precession
could play an important role, with possible consequences for the
explosion mechanism, r-process nucleosynthesis, or the measurable
neutrino
signal~\cite{Athar95,Totani96,Akhmedov97,Bruggen97,Nunokawa97b}.  The
downside of this richness of phenomena is that there are so many
unknown parameters (electromagnetic neutrino properties, masses,
mixing angles) as well as the unknown magnetic field strength and
distribution that it is difficult to come up with reliable limits or
requirements on neutrino properties.  The SN phenomenon is probably
too complicated to serve as a laboratory to pin down electromagnetic
neutrino properties, but it clearly is an environment where these
properties could have far-reaching consequences.

Resonant spin-flavor precessions can explain all solar neutrino
data~\cite{Akhmedov95,Guzzo98}, but require somewhat large toroidal
magnetic fields in the Sun since the neutrino magnetic (transition)
moments have to obey the globular-cluster limit of
Eq.~(\ref{eq:dipolelimit}).  The main original motivation for
magnetically induced oscillations was an apparent correlation between
the Homestake solar neutrino data and indicators of solar magnetic
activity.  Very recent re-analyses reveal that there is no significant
correlation with Sun spots~\cite{Walther97}, but also that the
hypothesis of a constant flux should be rejected with a significance
level of 0.1--6\%, depending on the test~\cite{Sturrock97}.  For
Majorana neutrinos, the spin-flavor precession amounts to transitions
between neutrinos and antineutrinos.  The observation of antineutrinos
from the Sun would be a diagnostic for this
effect~\cite{Barbieri91,Fiorentini97b,Pastor98}, and probably the only
convincing one.


\section{Conclusions}
\label{sec:conclusions}

As it stands, the most titillating question of neutrino physics no
longer is if these elusive particles have masses at all, but rather if
a fourth, hitherto unsuspected and otherwise noninteracting degree of
freedom exists to reconcile all current indications for neutrino
oscillations. If shockingly this were the case, the mass differences
suggested by LSND would imply that neutrinos are significant as a hot
dark matter component, corresponding to an eV-mass for one or two
flavors, which is what nowadays one means with a ``cosmologically
significant neutrino mass.'' Sterile neutrinos and a cosmological hot
dark matter component have become closely intertwined issues.

Oscillation experiments reveal only mass differences, leaving a common
offset from zero undetermined.  Even if LSND is right, the common mass
scale may exceed the indicated mass difference, and if LNSD is wrong
and sterile neutrinos do not exist, the sequential neutrinos could
still have nearly degenerate eV-masses and play a role for hot dark
matter. Fixing the common mass scale may soon become the major
challenge of neutrino physics.

There are few realistic opportunities to achieve this goal.  While
neutrinoless $\beta\beta$ decay experiments and precise tritium
endpoint $\beta$-spectra remain crucial, cosmology likely will play a
key role for this task.  The cosmological precision information
expected from the MAP and PLANCK microwave background missions and
from large-scale redshift surveys are in principle sensitive to
sub-0.1~eV masses.  Whether or not they will actually pin down such a
small mass remains to be seen, but surely they cannot ignore it as one
of about a dozen nontrivial cosmological parameters which are not
fixed by other data.

A direct kinematical mass limit from signal dispersion of a future
galactic supernova could get down to about 3~eV for $\nu_e$, probably
not good enough for the questions at hand. If high-energy neutrinos
from pulsed sources such as gamma-ray bursts are observed in upcoming
neutrino telescopes one may get down to much smaller masses.

The atmospheric neutrino anomaly requires a large mixing angle,
suggesting that all mixing angles in the neutrino sector could be
large, in blunt contrast to what is observed in the quark sector. A
large mixing angle between $\nu_e$ and other flavors radically changes
the interpretation of the SN~1987A neutrino signal and that from a
future galactic SN. Therefore, it is of paramount importance to
develop a better theoretical understanding of the neutrino spectra
formation in SNe to see if swapping flavors by oscillations indeed has
significant and observable effects. Apart from this important issue it
does not look as if neutrino oscillations had much to do with SN
physics itself, i.e.\ with the explosion mechanism, pulsar kicks, or
r-process nucleosynthesis, except perhaps if sterile neutrinos exist.

The large mixing angle implied by atmospheric neutrinos definitely
means that the neutrinos from ``cosmic beam dumps'' have a modified
flavor spectrum, presumably containing a large fraction of
$\nu_\tau$'s, which produce unique signatures in high-energy neutrino
telescopes.

Neutrino physics and neutrino astrophysics are at the cross roads. On
the one hand, it is now almost impossible to deny that neutrinos
oscillate and thus presumably have small masses. On the other hand,
unless a sterile neutrino truly exists, there is a sense that neutrino
masses are too small to be of very much cosmological or astrophysical
interest. Neutrino astrophysics could turn out to be more interesting
than one would have originally suspected, or more boring, depending on
whether sterile states exist or not.  

Either way, it may not be long until the neutrino mass and mixing
pattern has been reconstructed. The main beneficiary may be neutrino
astronomy. As we better understand the behavior of the neutrino beam
from distant sources, neutrino astronomy will return to its roots and
focus on the physics of the sources rather than worrying about the
behavior of the radiation. It may not be long until flavor
oscillations in neutrino astronomy are as commonplace a phenomenon as
the Faraday effect in radio astronomy!

\section*{Acknowledgments}

This work was supported, in part, by the Deutsche
Forschungsgemeinschaft under grant No.\ SFB-375.

\newpage


\end{document}